\documentclass[twocolumn]{aastex631}

\usepackage{booktabs}

\graphicspath{{./}{figures/}}

\begin{document}
\title{On the encounter between the GASP galaxy JO36 and the radio plume of GIN 049}
\shorttitle{On the encounter between the GASP galaxy JO36 and the radio plume of GIN 049}
\shortauthors{Ignesti et al.}

\correspondingauthor{Alessandro Ignesti}
\email{alessandro.ignesti@inaf.it}

\author[0000-0003-1581-0092]{Alessandro Ignesti}\affiliation{INAF-Padova Astronomical Observatory, Vicolo dell’Osservatorio 5, I-35122 Padova, Italy}

\author[0000-0003-4120-9970]{Marisa Brienza}\affiliation{INAF – Osservatorio di Astrofisica e Scienza dello Spazio di Bologna, Via P. Gobetti 93/3, 40129 Bologna, Italy}\affiliation{Dipartimento di Fisica e Astronomia, Universiti di Bologna, via P. Gobetti 93/2, 40129 Bologna, Italy}

\author[0000-0003-0980-1499]{Benedetta Vulcani}\affiliation{INAF-Padova Astronomical Observatory, Vicolo dell’Osservatorio 5, I-35122 Padova, Italy}

\author[0000-0001-8751-8360]{Bianca M. Poggianti}\affiliation{INAF-Padova Astronomical Observatory, Vicolo dell’Osservatorio 5, I-35122 Padova, Italy}

\author[0000-0002-5655-6054]{Antonino Marasco}\affiliation{INAF-Padova Astronomical Observatory, Vicolo dell’Osservatorio 5, I-35122 Padova, Italy}

\author[0000-0001-5303-6830]{Rory Smith}\affiliation{Universidad T\`ecnica Federico Santa Mar\`ia
Avda. Vicu\~na Mackenna 3939, Oficina A035, San Joaqu\`in, Santiago, Chile}

\author[0000-0003-4223-1117]{Martin J. Hardcastle}\affiliation{Centre for Astrophysics Research, Department of Physics, Astronomy and Mathematics, University of Hertfordshire, College Lane, Hatfield AL10 9AB UK}

\author[0000-0002-9325-1567]{Andrea Botteon}\affiliation{INAF, Istituto di Radioastronomia di Bologna, via Piero Gobetti 101, 40129 Bologna, Italy}

\author[0000-0002-0692-0911]{Ian D. Roberts}\affiliation{Leiden Observatory, Leiden University, PO Box 9513, 2300 RA Leiden, The Netherlands}

\author[0000-0002-7042-1965]{Jacopo Fritz}\affiliation{Instituto de Radioastronomía y Astrofísica, UNAM, Campus Morelia, A.P. 3-72, C.P. 58089, Mexico}

\author[0000-0001-9143-6026]{Rosita Paladino}\affiliation{INAF, Istituto di Radioastronomia di Bologna, via Piero Gobetti 101, 40129 Bologna, Italy}

\author[0000-0002-0843-3009]{Myriam Gitti}\affiliation{Dipartimento di Fisica e Astronomia, Università di Bologna, via Piero Gobetti 93/2, 40129 Bologna, Italy}\affiliation{INAF, Istituto di Radioastronomia di Bologna, via Piero Gobetti 101, 40129 Bologna, Italy}

\author[0000-0001-5840-9835]{Anna Wolter}\affiliation{INAF - Osservatorio Astronomico di Brera, via Brera, 28, 20121, Milano, Italy}

\author[0000-0002-8238-9210]{Neven Tomi\v{c}i\'{c}}\affiliation{Department of Physics, Faculty of Science, University of Zagreb, Bijenicka 32, 10 000 Zagreb, Croatia}

\author[0000-0003-3255-3139]{Sean McGee}\affiliation{School of Physics and Astronomy, University of Birmingham, Birmingham B15 2TT, United Kingdom}

\author[0000-0002-1688-482X]{Alessia Moretti}\affiliation{INAF-Padova Astronomical Observatory, Vicolo dell’Osservatorio 5, I-35122 Padova, Italy}

\author[0000-0002-7296-9780]{Marco Gullieuszik}\affiliation{INAF-Padova Astronomical Observatory, Vicolo dell’Osservatorio 5, I-35122 Padova, Italy}

\author[0000-0003-2792-1793]{Alexander Drabent}\affiliation{Th\"uringer Landessternwarte, Sternwarte 5, D-07778 Tautenburg, Germany }

\begin{abstract}

We report on the serendipitous discovery of an unprecedented interaction between the radio lobe of a radio galaxy and a spiral galaxy. The discovery was made thanks to LOFAR observations at 144 MHz of the galaxy cluster Abell 160 ($z=0.04317$) provided by the LOFAR Two-metre Sky Survey. The new low-frequency observations revealed that one of the radio plumes of the central galaxy GIN 049 overlaps with the spiral galaxy JO36. Previous studies carried out with MUSE revealed that the warm ionized gas in the disk of JO36, traced by the H$\alpha$ emission, is severely truncated with respect to the stellar disk.  We further explore this unique system by including new uGMRT observations at 675 MHz to map the spectral index. The emerging scenario is that JO36 has interacted with the radio plume in the past 200-500 Myr.  The encounter resulted in a positive feedback event for JO36 in the form of a star formation rate burst of $\sim14$ $M_\odot$ yr$^{-1}$. In turn, the galaxy passage left a trace in the radio-old plasma by re-shaping the old relativistic plasma via magnetic draping. 
\end{abstract}



\section{Introduction} \label{sec:intro}
Galaxy clusters are astrophysical laboratories to study baryon physics and galaxy evolution. 
\begin{figure*}
\includegraphics[width=\linewidth]{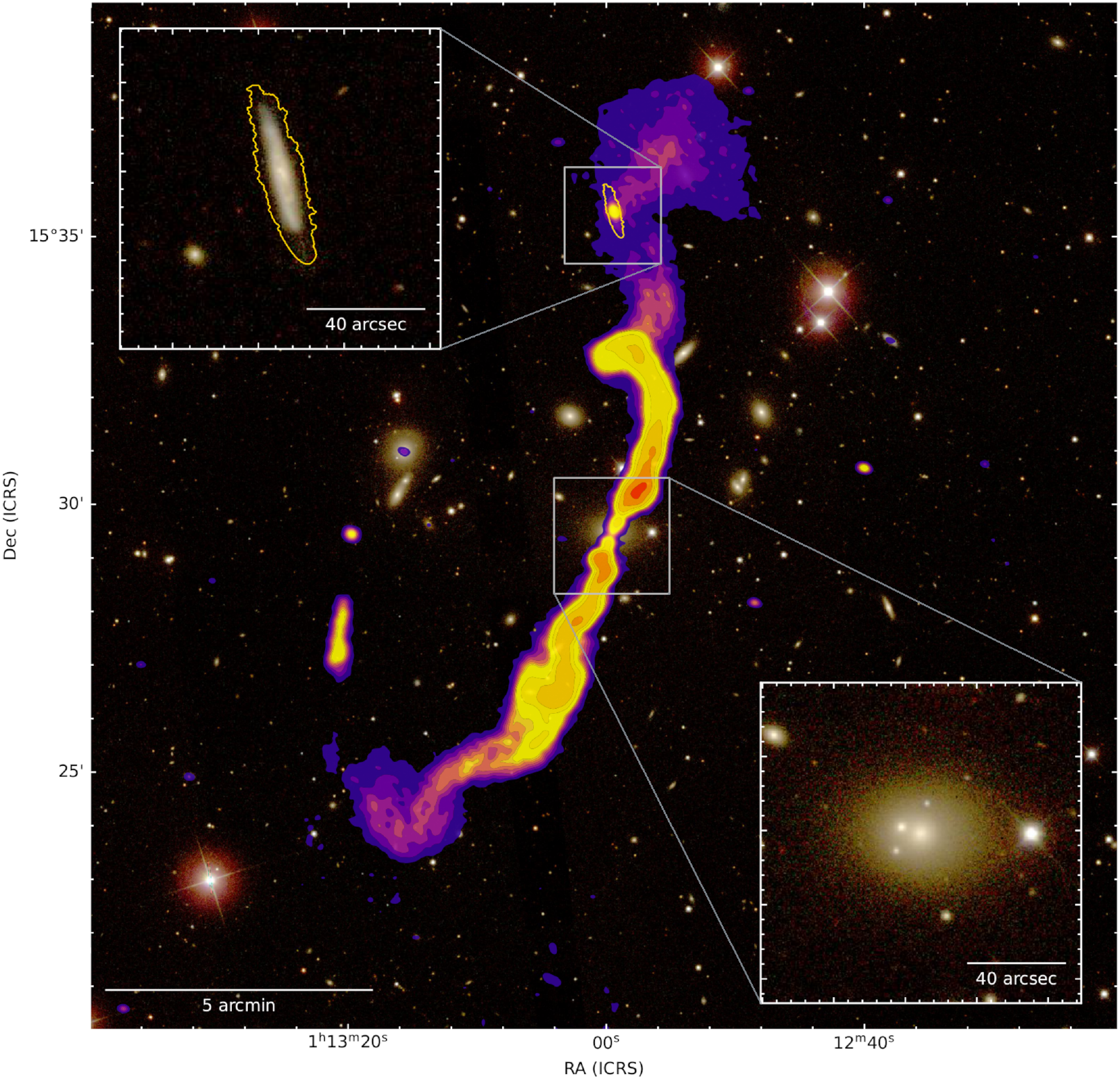}
\label{panel}
\caption{SDSS \texttt{gri} image of A160 overlaid with the 3, 6, 9, 12, 15, 18, 21, 24, 48, 80, 160, 320, 550$\times$RMS levels of the 144 MHz radio emission (RMS=127 $\mu$Jy beam$^{-1}$, resolution=$10.6\times5.2$ arcsec$^2$) and stellar disk of JO36 (gold contour). In the top-left and bottom-right insert we report the zoom-in on, respectively, JO36 and GIN-049.}
\end{figure*}
It has been widely shown in the past decades that the extreme environments of galaxy clusters can strongly influence the properties of galaxies within. This `environmental processing' can take place via either gravitational or hydrodynamics effects, and can result in galaxies with a plethora of features raging from disturbed/truncated ionized plasma disks to long tails of debris \citep[e.g.,][and references therein]{Moore1996, vanGorkom_2004,Boselli_2022,Cortese_2021}. At radio wavelengths, in galaxy clusters, we can observe large-scale radio sources hosted by giant elliptical galaxies. They are powered by the relativistic plasma ejected by their central Active Galactic Nuclei (AGN), where the relativistic electrons emit at radio wavelengths via synchrotron emission due to the presence of magnetic fields and their properties are driven also by their interplay with the surrounding intracluster medium (ICM) \citep[e.g.,][]{Fanaroff-Riley_1974,O'Dea_1985, Burns_1990, Hardcastle_2020} 

Here we report on the serendipitous discovery of an interaction between a ram-pressure-stripped galaxy, and the radio lobe of a radio galaxy. The discovery was made possible thanks to radio continuum observations of the galaxy cluster Abell 160 (hereafter A160) at 144 MHz carried out with the Low-Frequency Array \citep[LOFAR,][]{vanHaarlem_2013}. The new low-frequency observation, which we showcase in Figure \ref{panel}, reveals that the northern radio plume of the central galaxy GIN 049 encompasses the spiral galaxy JO36. This discovery emerged in the context of a series of LOFAR observations of ram pressure-stripped galaxies from the GASP sample \citep[GAs Stripping Phenomena in galaxies with MUSE][]{Poggianti_2016}. 

JO36-GIN 049 is a unique laboratory to study the properties of relativistic plasma under the circumstances of a stellar disk-radio lobe interaction. In order to pave the way for future studies, in this work, along with presenting the observations, we aim to characterize this encounter in terms of feedback and implications for both JO36 and the radio lobe. 
In the remainder of the paper, we present a multi-wavelength study that involves new radio continuum observations from LOFAR and upgraded Giant Metrewave Radio Telescope (uGMRT), and optical Multi Unit Spectroscopic Explorer (MUSE) observations. The manuscript is structured as follows. In Section 2 we present the previous studies on the two galaxies and the data preparation. The new results are presented in Section 3, including the properties of a puzzling radio source, the `Mandolin', which we observe nearby GIN 049, and a series of new, high-resolution spectral index maps. Finally, in Section 4 we present a series of analyses aimed at characterizing the outcomes of the encounter and draw our conclusions. We also present a novel method to estimate the expected truncation radius induced by the ram pressure, described in detail in Appendix \ref{appendix_truncation}. 

Throughout the paper, we adopt a $\Lambda$CDM cosmology with $\Omega_{\Lambda}=0.7$, $\Omega_{\rm m}=0.3$, and $H_0=70$ km s$^{-1}$ Mpc$^{-1}$, which yields $1''=0.851$ kpc\footnote{\url{https://www.astro.ucla.edu/\%7Ewright/CosmoCalc.html}} at the cluster redshift \citep[$z=0.04317$,][]{Gullieuszik_2020}. We describe the radio synchrotron spectrum as $S\propto\nu^\alpha$, where $S$ is the flux density, $\nu$ is the frequency, and $\alpha$ is the spectral index.
\section{Target description and Data analysis}
\subsection{JO36}
\label{jo36}
\begin{figure}
    
    \includegraphics[width=\linewidth]{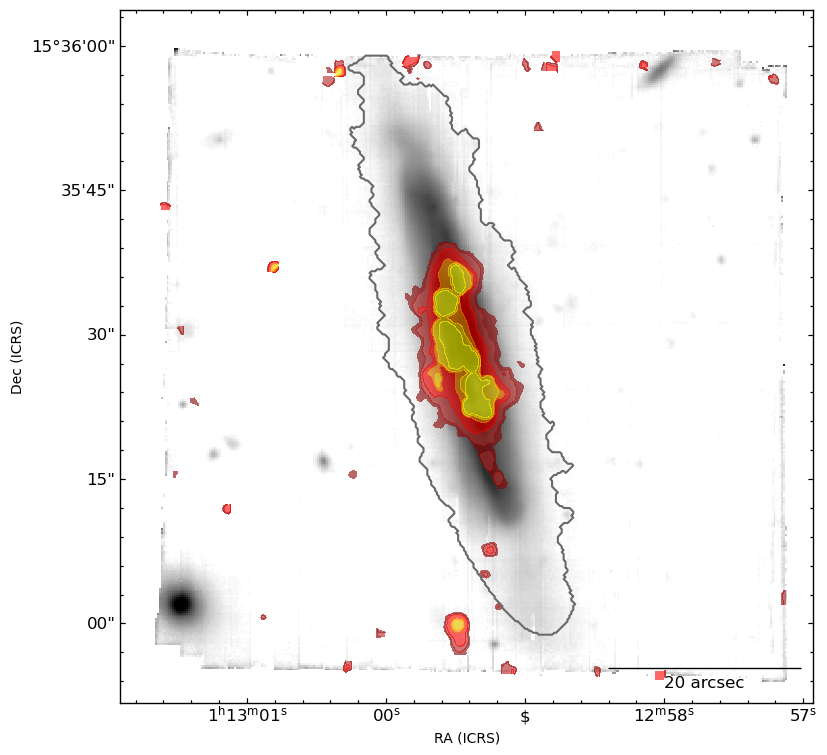}
    \caption{MUSE total intensity image of JO36 with the contours of the stellar disk \citep[black contour, from][]{Gullieuszik_2020} and the emission-only H$\alpha$ emission (red-to-yellow filled).}
    \label{halpha}
\end{figure}

JO36 (RA 18.2476, DEC 15.5914, a.k.a 2MFGC 00903) is a galaxy in A160 with redshift $z$=0.0407 \citep[][]{Fritz2017}, and has been observed in the MUSE survey led by the GASP collaboration \citep[][]{Poggianti2017}. This galaxy, which is shown in Figure \ref{halpha}, displays truncated H$\alpha$ emission suggesting an advanced stripping phase, but no evidence of stripped tails of ionized gas such as those commonly encountered in the so-called jellyfish galaxies \citep[][]{Fumagalli2014, Smith2010, Ebeling2014,Poggianti2017}. The stripping direction of JO36 is unclear due to contrasting features, such as the presence of ionized gas in the southern part of the galaxy, which indicates a velocity component toward the north, and the distorted shape of the gas rotational axis, which suggests instead a velocity component toward the south \citep[][]{Fritz2017}. JO36 currently hosts a moderately high star formation rate (SFR) of $6\pm1$ $M_\odot$ yr$^{-1}$ \citep[][]{Vulcani2018}, with evidence of a past intense star-forming episode that happened in the past 200-500 Myr \citep[][]{Fritz2017}. Finally, strong central X-ray emission may suggest the presence of an obscured AGN \citep[][]{Fritz2017, Peluso_2021}. \citet[][]{Fritz2017} concluded that this peculiar galaxy could be the result of a combination of ram pressure and tidal effects in action. 

\subsection{GIN 049}
\label{gin049}
GIN 049 (RA 18.2482, DEC 15.4912, a.k.a. B0110+152) is a dumbbell galaxy with 4 main optical cores and is the Brightest Cluster Galaxy (BCG) of A160. At radio wavelengths, it is classified as wide-angle tail \citep[][]{O'Donoghue_1990} FRI radio galaxy \citep[][]{Fanaroff-Riley_1974}. It was among the first radio galaxies studied in detail with VLA and WSRT observations at 6 and 20 cm \citep[e.g.,][]{Wirth_1982,Fanti_1983, Giovannini_1987, O'Donoghue_1990, Parma_1991} covering both the central emission and the jets. The radio source appears highly collimated, with a bend toward East in the northern jet, and an S-shape in the southern one. The peculiar morphology has been attributed to gravitational interactions between the cores \citep[][]{Wirth_1982,Parma_1991}. Due to the lack of lobes in these observations, it would be an example of a `plumed' FRI galaxy. Polarization studies revealed that the magnetic field is aligned with the jet direction \citep[][]{Giovannini_1987}, although the polarized emission intensity appears to be less extended in the northern jet than in the southern one. Finally, the 1.4-5 GHz spectral index steepens from -0.5 to -1.5 going from the center to the plumes \citep[][]{O'Donoghue_1990}. Combined radio and X-ray studies revealed a potential connection between the radio morphology and the ICM distribution, where the presence of substructures in the ICM has driven the evolution of the radio jets. Specifically, \citet[][]{Burns_1994} observed unusual elongations or clumps in the central X-ray emission of A160 observed by ROSAT, thus suggesting that the radio galaxy could have developed after a recent merger between  substructures. Subsequently, \citet[][]{Jetha_2005} reported that the location where the jet rapidly expands into the plume corresponds to a steep ICM temperature gradient that separates the cluster cool core from the rest of the ICM. 

\subsection{Radio observations and data reduction}
 Images at 120-168 MHz were provided by the LOFAR Two-metre Sky Survey \citep[LoTSS, ][]{Shimwell_2017,Shimwell_2019,Shimwell_2022}, pointing P018+16. For details on the observational strategy and calibration of the LoTSS data we refer to \citet[][]{Shimwell_2022}. Following the LoTSS procedure, the dataset is calibrated by using the direction-dependent calibration and imaging pipeline {\scshape ddf-pipeline}\footnote{\url{ https://github.com/mhardcastle/ddf-pipeline}} v. 2.2. This pipeline was developed by the LOFAR Surveys Key Science Project and it corrects for ionospheric and beam model errors in the data. The latest version of the pipeline is described by \citet[][]{Tasse_2021}. The entire data processing procedure makes use of {\scshape prefactor}  \citep[][]{vanWeeren_2016,Williams_2016,deGasperin_2019}, {\scshape killMS} \citep[][]{Tasse_2014,Smirnov_2015} and {\scshape DDFacet} \citep[][]{Tasse_2018}. The observation has been further processed to improve the calibration in a $\sim0.8\times0.8$ degree$^2$ region of the LOFAR field-of-view containing the target of interest, employing the ``extraction and self-calibration" procedure described by \citet[][]{vanweeren_2021}. 

GIN-049 was observed with the uGMRT in band 4 (550--950 MHz) for 6h on February 18, 2023 (project code: 43\_009, PI: A. Ignesti). The observations include the flux density calibrators J0117+143 and 3C48. Data were recorded in 4096 frequency channels with an integration time of 5.3~s using both the narrow-band (bandwidth of 33.3 MHz) and wide-band (bandwidth of 400 MHz) backends. Data are processed with the Source Peeling and Atmospheric Modeling (SPAM\footnote{\url{http://www.intema.nl/doku.php?id=huibintemaspampipeline}}) package \citep[][]{Intema_2009}, which performs calibration of the flux density scale, correction for the bandpass, data averaging and flagging, and direction-dependent self-calibration. We split the dataset into 8 sub-bands and then proceed with the calibration separately. In order to improve the calibration, we use a sky model produced from the narrow-band observation as an input for the self-calibration. In the end, we drop the three slices above 800 MHz, due to severe RFI. In order to further improve the calibration of the 700-750 MHz sub-band, we perform additional self-calibration by using as reference a model of the target created with the CASA software v 4.3.1 \citep[][]{CASA_2022}. 

We produce the images by using {\scshape WSClean}  v2.10.1  \citep[][]{Offringa_2014} and by making use of the HOTCAT cluster \citep[][]{Taffoni_2020,Bertocco_2020}. The final images produced from the LOFAR and uGMRT observations, which have a central frequency of, respectively, 144 and 675 MHz, are obtained by using the `joined-channel deconvolution'\footnote{\url{https://wsclean.readthedocs.io/en/latest/wideband_deconvolution.html}} of {\scshape WSClean} with 10 channels. We adopt different combinations of \texttt{ROBUST} \citep[][]{Briggs_1994}, UV taper, and UV cut to map the radio emission at different scales and angular resolution, which we show in Figures \ref{comparison} and \ref{mistero}. The angular resolution, the RMS, and the clean parameters of each map are summarized in Table \ref{table_images}.

\begin{table*}[]
\centering
\begin{tabular}{lccccccc}
\toprule
\toprule
Figure & Frequency& Beam & RMS & \multicolumn{3}{c}{Clean parameters} \\
& [MHz] & [arcsec$^2$] &[$\mu$Jy beam$^{-1}$]&\texttt{ROBUST}&\texttt{UVMIN}&\texttt{UVTAPER}\\
\midrule
\ref{panel}                & 144   &       10.6$\times$5.2        &      127        &       \texttt{-0.5}&-&-\\
\midrule
\ref{comparison}, top-left      & 144   &8.6$\times$4.6&      110        &  \texttt{-0.7}&-&-  \\
\ref{comparison}, top-right     & 675   &  13.6$\times$8.2 &    110          &  \texttt{-0.5}&  -&\texttt{9 arcsec} \\
\ref{comparison}, bottom-left   & 144   &    11.3$\times$5.2      &    159   &  \texttt{-0.7}&  \texttt{700} $\lambda$&- \\
\ref{comparison}, bottom-right & 675   &  5.4$\times$2.8   &      40        &    \texttt{-0.7}&-&-\\
\midrule
\ref{mistero} & 144   & 17.7$\times$11.4  &      172  &    \texttt{-0.5}&-&  \texttt{10 arcsec}    \\
\midrule
\ref{alfa-image}, top           & 144   &   9$\times$9   &     135   &  \texttt{-0.7}&  \texttt{300} $\lambda$&-\\
\ref{alfa-image}, top           & 675   &    9$\times$9  &      81      & \texttt{-0.7}&  \texttt{300} $\lambda$&-\\
\ref{alfa-image}, center        & 144   &    17$\times$17 &       232       & \texttt{-0.7}&  \texttt{300} $\lambda$& \texttt{10 arcsec} \\
\ref{alfa-image}, center        & 675   &  17$\times$17&          194  &\texttt{-0.7}&  \texttt{300} $\lambda$& \texttt{10 arcsec}\\
\ref{alfa-image}, bottom        & 144   &   30$\times$30 &       390     &\texttt{-0.7}&  \texttt{300} $\lambda$& \texttt{20 arcsec} \\
\ref{alfa-image}, bottom        & 675   &  30$\times$30&        428  & \texttt{-0.7}&  \texttt{300} $\lambda$& \texttt{20 arcsec} \\ \midrule
\end{tabular}
\caption{\label{table_images}Summary of radio images produced from the LOFAR and uGRMT observations at 144 and 675 MHz, respectively, and presented in this work. From left to right: figure, central frequency, angular resolution, RMS, and relevant parameters used in \texttt{WSCLEAN}. For Figure \ref{alfa-image} we report the parameters of the images used to compute the spectral index maps.}
\end{table*}

We combine the 144 and 675 MHz images to map the spectral index over the radio source. First, we produce images at 144 and 675 MHz with a lower UV cut of 300$\lambda$ ($\sim13$ arcmin), which is necessary to match the largest angular scales mapped by the two observations, and different UV tapering to produce images with increasing beam size and, hence, sensitivity to extended, faint emission. Then we convolve them with a Gaussian beam with different resolutions (9, 17, and 30 arcsec) to match the angular resolution, correct the astrometry with a precision below $\sim1''$ to improve the overlap between the two images and combine them to compute the spectral index maps as:
\begin{equation}
\alpha=\frac{\log\left(\frac{S_{144}}{S_{675}}\right)}{\log\left(\frac{144}{675}\right)}\pm\frac{1}{\log\left(\frac{144}{675}\right)}\sqrt{\left(\frac{\sigma_{S,144}}{S_{144}}\right)^2+\left(\frac{\sigma_{S,675}}{S_{675}}\right)^2}\text{,}
    \label{alfa}
\end{equation}
where $S$ and $\sigma_S$ are, respectively, the flux density and the corresponding uncertainties in each pixel. We assume $\sigma_S$ to correspond to the RMS of our images (see Table \ref{table_images}). We set a lower surface brightness threshold of 2$\times$RMS (determined individually in each band, see Table \ref{table_images}), and assume the RMS as the uncertainty on each pixel. The resulting images are shown in Figure \ref{alfa-image}.
\subsection{Ancillary optical data}
We combine SDSS \citep[][]{Donald_2000} images\footnote{From \url{https://dr12.sdss.org/mosaics/}} in the {\tt g-}, {\tt r-} and {\tt i}-filter to produce color images of the cluster (Figure \ref{panel}). We make use of the previous results obtained by the GASP team from the MUSE optical observations \citep[][]{Poggianti2017,Fritz2017}. Specifically, we use the emission-only H$\alpha$ emission map (Figure \ref{halpha}, angular resolution $\sim$1 arcsec$^2$) corrected for extinction due to the dust in the Milky Way, using the value estimated at the galaxy position \citep[][]{Schlafly2011} and assuming the extinction law from \citet[][]{Cardelli1989}. The emission-only data cube is obtained by subtracting the stellar-only component of each spectrum derived with our spectrophotometric code SINOPSIS \citep[SImulatiNg OPtical Spectra wIth Stellar population models\footnote{\url{https://www.irya.unam.mx/gente/j.fritz/JFhp/SINOPSIS.html}},][]{Fritz2017}. This code searches the combination of Single Stellar Population (SSP) model spectra that best fits the equivalent widths of the main lines in absorption and emission and the continuum at various wavelengths and provides spatially resolved estimates of a number of stellar population properties, such as stellar masses and star formation histories. SINOPSIS provides also information on the star formation history (SFH) in the form of average SFR in twelve logarithmically-spaced age bins \citep{Fritz2017}. \\

\section{Results}
\subsection{Radio images}
\label{radio_res}
\begin{figure*}
\centering

\includegraphics[width=.87\textwidth]{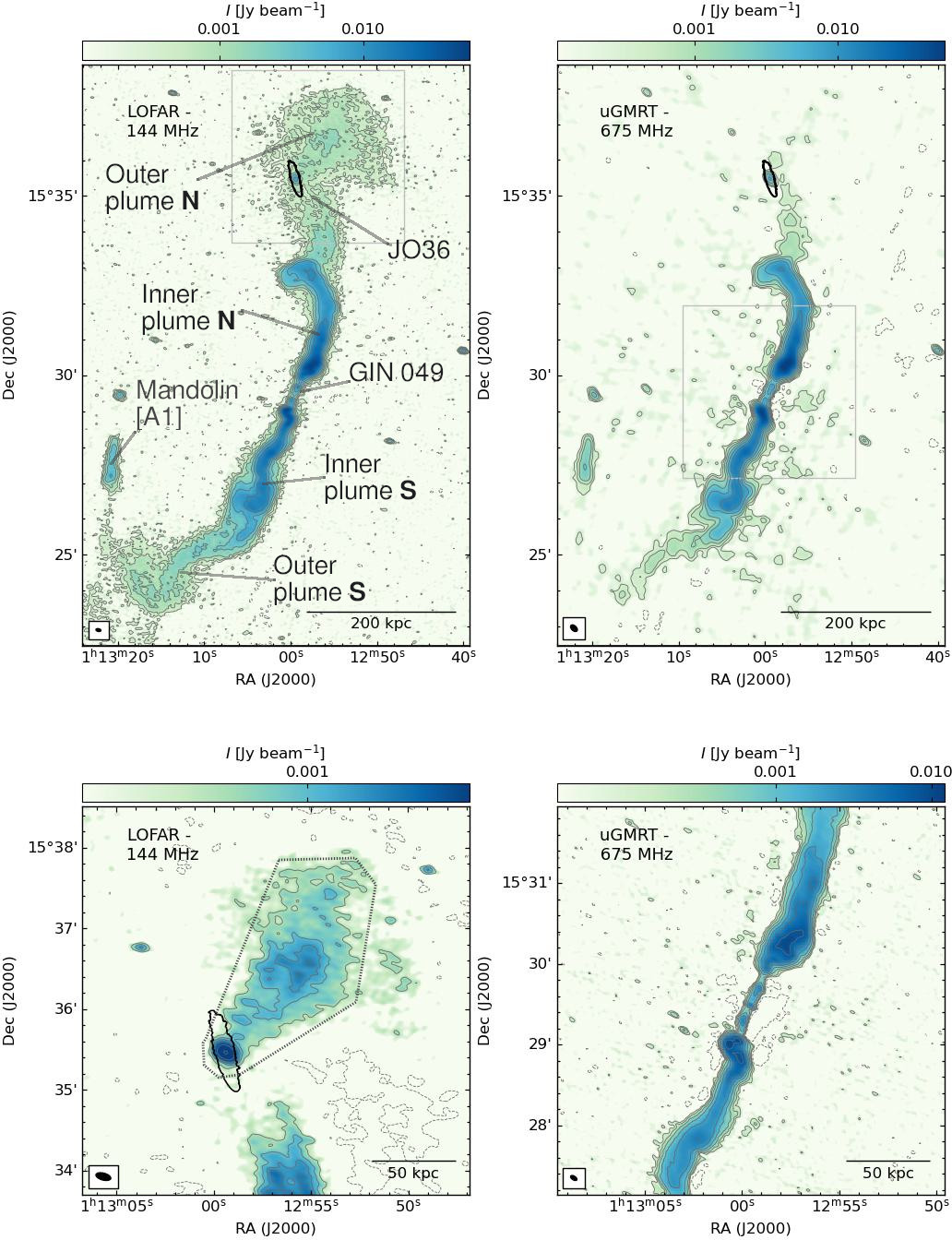}
\caption{\label{comparison}Radio continuum images of the JO36-GIN 049 system. Top-left: 144 MHz (RMS=110 $\mu$Jy beam$^{-1}$, resolution=$8.6\times4.6$ arcsec$^2$); Top-right: 675 MHz (RMS=110 $\mu$Jy beam$^{-1}$, resolution=$13.6\times8.2$ arcsec$^2$); Bottom-left: 144 MHz (RMS=159 $\mu$Jy beam$^{-1}$, resolution=$11.3\times5.2$ arcsec$^2$). The grey, dashed polygon defines the `extended region' (see Section \ref{radio_res}); Bottom- right: 675 MHz  (bottom right, RMS=40 $\mu$Jy beam$^{-1}$, resolution=$5.4\times2.8$ arcsec$^2$). The relevant parameters used to produce the images with \texttt{WSCLEAN} are reported in Table \ref{table_images}. We show the -3, 3, 6, 12, 24, 48, 96$\times$RMS levels (grey), and the contour of the stellar disk of JO36 (black). The silver rectangular regions in the top panels indicate the FOV of the lower panels.}
\end{figure*}

The radio emission of GIN 049 obtained with the new images is more extended than that presented in the literature \citep[$\sim8$ arcmin][]{Giovannini_1987}. Specifically, in the new LOFAR image (Figure \ref{panel}) we observe two new components, both toward the south and the north, that extend the radio source projected angular size at 144 MHz to $\sim15.5$ arcmin, corresponding to $\sim 744$ kpc at the cluster redshift. Having a physical size larger than 700 kpc, GIN 049 can thus be classified as a `giant' radio galaxy \citep[][]{Dabhade_2020}. In the following, we refer to the emission within $\sim200$ arcsec ($\sim$160 kpc) from the center, already studied in the literature, as `inner' plumes, and the new extension detected by LOFAR as `outer' plumes (see Figure \ref{comparison}, top-left panel). At 675 MHz we detect only part of the outer plumes (Figure \ref{comparison}, top-right panel). In the high-resolution image at 675 MHz map of the central region (Figure \ref{comparison}, bottom-right panel) we clearly observe the central core and the two very collimated jets, that eventually expand in the plumes. For reference, in the maps presented in Figure \ref{comparison} (top panels), we measure within the 3$\times$RMS contour a flux density of $5.51\pm0.55$\footnote{We include the emission coincident with JO36, that is $\sim18$ mJy (0.3$\%$).} and $1.62\pm0.16$ Jy at 144 and 675 MHz, respectively. The uncertainty on the integrated flux density $S$ is computed as $\sigma_S=\sqrt{(f\cdot S)^2+(\text{RMS}\sqrt{A/A_{\text{beam}}})^2}$, where $f=0.1$ is the flux scale error and $A/A_{\text{beam}}$ is the ratio between the source and the beam areas. These values yield a radio luminosity of 24.1$\times10^{24}$ and 7.1$\times10^{24}$ W Hz$^{-1}$ at the cluster redshift, which is consistent with the FRI classification. The radio luminosity is computed as $L=4\pi D_l^2S(1+z)^{-\alpha-1}$ by using the cluster luminosity distance $D_l=191$ Mpc and a spectral index $\alpha=-0.8$ derived from the integrated flux densities by following Equation \ref{alfa}. The latter is consistent with the previous studies (see Section 2.2) and, in general, with the average spectral index of active radio sources \citep[e.g.,][]{Parma_1999}.

The morphology of the inner plumes agrees with previous studies (see Section \ref{gin049}). The northern one appears to be mostly aligned toward N-W before turning toward East, whereas the southern one is more S-shaped (see Figure \ref{comparison}, bottom right panel). LOFAR observations show that the serpentine morphology continues in the southern plume that gets progressively wider with distance, which is expected from the relativistic plasma expansion at increasing distances from the galaxy. The northern plume begins with a sharp decline in brightness and a re-alignment along the N-W direction, followed by the final expansion in which it embeds JO36.

At 144 MHz JO36 appears as a compact source embedded in the northern outer plume. In the bottom left panel of Figure \ref{comparison}, we show the result of applying a lower UV-cut of 700 $\lambda$ to remove the emission on scales larger than $\sim1$ arcmin, which is roughly the outer plume angular size. We clearly observe the emission associated with the stellar disk, which seems to overlap with the H$\alpha$ emission (see Figure \ref{halpha}), and, in addition, a `tail-like' substructure that, starting from JO36, is elongated toward N-W for $\sim1.9$ arcmin ($\sim140$ kpc). We also observe a clear gap in the emission that separates JO36 from the northern plume. At 675 MHz (Figure \ref{panel}, top-right panel) we detect the emission from the stellar disk but the extended component detection is less striking and the `tail' appears, albeit narrower, only in the smoothed images (see Section \ref{analisi_alfa}, bottom panels). The real extent of JO36 radio emission is unclear, thus we measure the flux densities in two regions, which are the stellar disk defined in \citet[][]{Gullieuszik_2020} (black contour in Figure \ref{halpha}), and the region defined by the dashed polygon shown in Figure \ref{comparison} (bottom-left panel) which hereafter we refer to as `extended region'. Within the stellar disk, we measure a flux density of $18.4\pm1.8$ and $7.5\pm0.8$ mJy at 144 and 675 MHz, respectively. The corresponding spectral index is $\alpha=-0.6\pm0.1$ which implies, at the cluster redshift, luminosities of 8.0 and 3.2$\times10^{22}$ W Hz$^{-1}$. In the extended region, we measure a 144 MHz flux density of $124.1\pm12$ mJy within the $3\times$RMS contour, which implies a luminosity of $5.4\times10^{23}$ W Hz$^{-1}$.%
\subsubsection{The `Mandolin'}
\begin{figure}
    \centering
    \includegraphics[width=\linewidth]{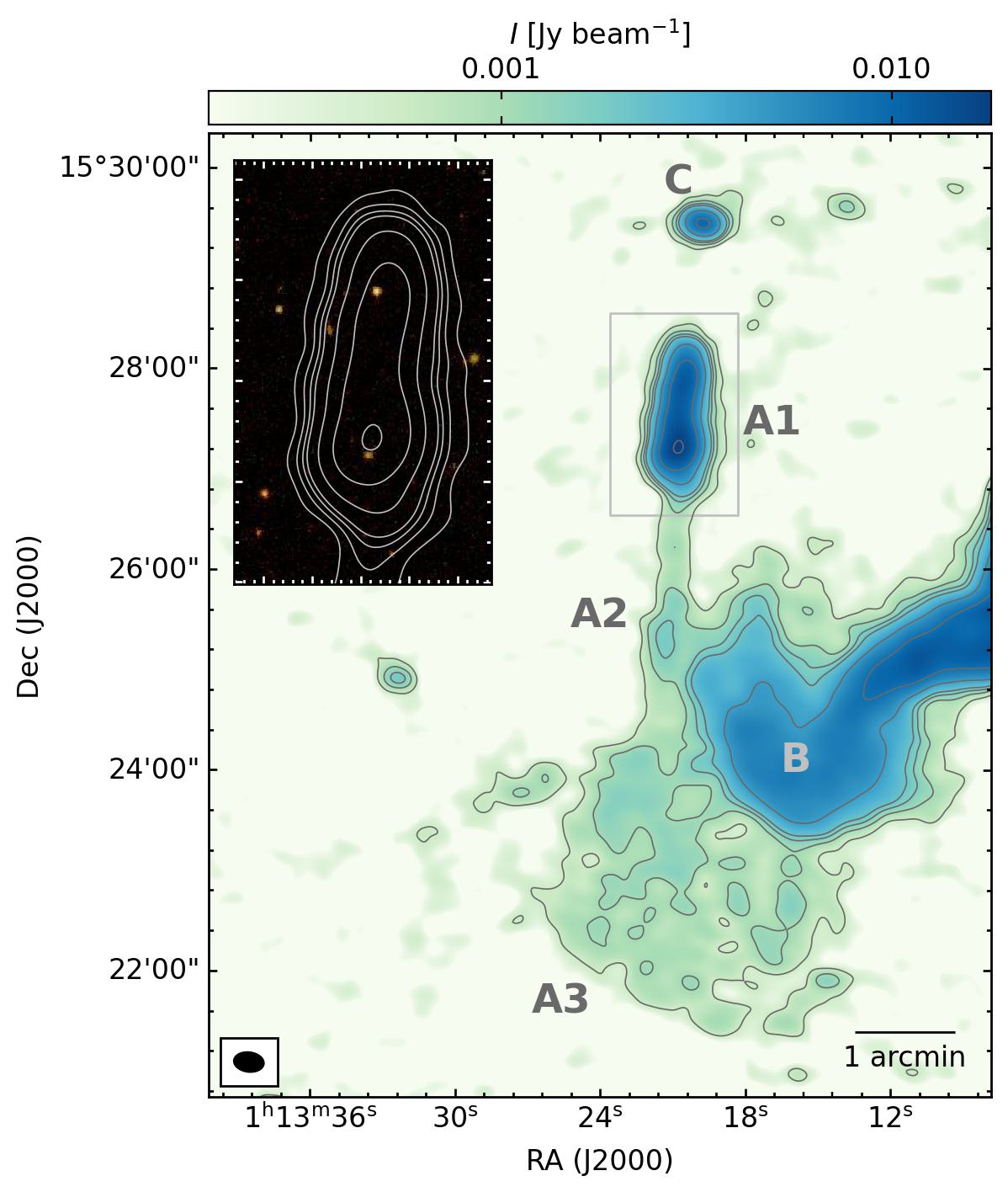}
    \caption{Radio continuum image at 144 MHz of the `Mandolin' radio source with the contours at the 3, 6, 12, 24, 48$\times$RMS (RMS=172 $\mu$Jy beam$^{-1}$, resolution=$17.7\times11.4$ arcsec$^2$). The labels point out the Mandolin (where headstock, neck, and body are, respectively, A1, A2, and A3), GIN 049 southern lobe (B), and WINGS J011319.77+152927.4 (C). In the top-left insert, we show the SDSS optical image of the headstock region (silver box).}
    \label{mistero}
\end{figure}
We report on the discovery of a second, extremely puzzling source detected near to GIN 049 that we call the `Mandolin'. The radio source, shown in Figure \ref{mistero}, is composed of an elongated, bright component toward the north, the `headstock' (region A1 in Fig.\,\ref{mistero}), an elongated ridge, the `neck' (A2), along the north-south direction, that ends in a low-brightness lobe, the `body' (A3). The two latter components are detected only at 144 MHz. Sitting directly above the `headstock' (and remarkably aligned with the `neck'), we observe WINGS J011319.77+152927.4, which is a spiral star-forming galaxy (source C in Figure \ref{mistero}). The body appears to be connected to the southern lobe of GIN 049, but we cannot discriminate if the connection is real, so that A3 is a continuation of the southern plume, or a projection effect. The integrated flux density at 144 MHz, excluding GIN 049 southern plume emission (which we identify as the emission encompassed by the $12\times$RMS contour), is $213.3\pm21$ mJy, of which 106.2 mJy (49.8$\%$) resides in the `headstock'. This latter component is detected also at 675 MHz (see Figure \ref{comparison}, top right panel), at which we measure a flux density of $25.3\pm2.6$ mJy. The resulting spectral index is $\alpha_{144}^{675}=-0.9\pm0.1$. The headstock is detected also at 1.4 GHz in the NRAO VLA Sky Survey images \citep[][]{Condon_1998}, from which we measure a flux density of $\sim8.5$ mJy. This entails a spectral index $\alpha_{675}^{1400}\simeq-1.5$ between 675 MHz and 1.4 GHz, so, that the headstock has the characteristic curved synchrotron spectrum of fossil radio plasma \citep[e.g.,][]{Mandal_2020, Ignesti_2020b, Brienza_2022}.

We can only speculate on the origin of the Mandolin. We cannot identify any evident optical counterpart, so it would appear to be a diffuse source. The two optical sources encompassed in A1 are WISEA J011320.67+152756.4 (north) and WISEA J011320.84+152707.7 (south), for which the nature is unclear and there is no redshift in the literature\footnote{\url{https://ned.ipac.caltech.edu/}}. It would be possible that the headstock (A1) is the blending of their radio emission. Assuming that the Mandolin is physically located inside A160, its angular size, $\sim6.7$ arcmin, would correspond to $\sim320$ kpc, which would not be unusual for a remnant radio source, especially for a so-called `radio phoenix' \citep[][]{VanWeeren_2019, Mandal_2020}. Thus it might be possible that the Mandolin is an old radio source composed of fossil plasma previously injected by an AGN. In this scenario, the most likely host would be GIN 049, in which case the Mandolin could be just the `tip of the iceberg' of a larger, underlying structure of fossil radio plasma ejected by a previous AGN phase \citep[similar to the case of Nest200047 radio source, see][]{Brienza_2021}. In this scenario, A3 could be the faint continuation of the southern lobe (B). We observe that A3 roughly coincides with the ICM substructures discussed in \citet[][]{Burns_1994}, thus it might be possible that the ICM might have played a role in the origin of this radio source, either via compression or re-acceleration of fossil radio plasma.

\subsection{Spectral index map}
\label{analisi_alfa}
\begin{figure*}
\centering
    \includegraphics[width=\linewidth]{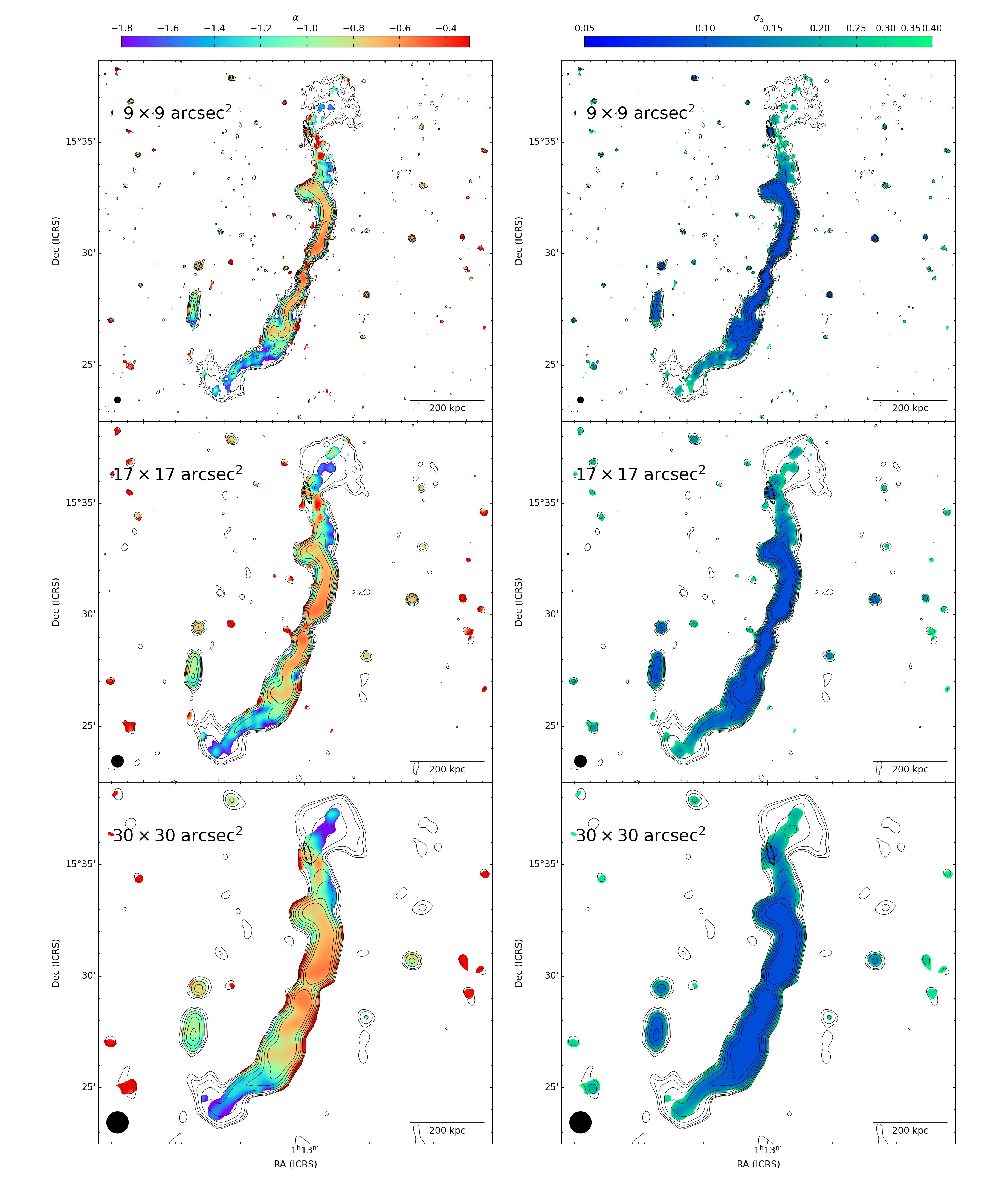}
    \caption{Spectral index maps between 144 and 675 MHz (left), and corresponding uncertainties (right). The angular resolution, from top to bottom, is 9, 17, and 30 arcsec. For each angular resolution, we report the 3, 6, 12, 24, 48, 96, 192$\times$RMS contours of the corresponding LOFAR image.}
    \label{alfa-image}
\end{figure*}

The spectral index maps (Figure \ref{alfa-image}) show a spectral steepening with increasing distance from GIN 049, as expected in this kind of radio source where the relativistic particles are accelerated at the base of the plumes and then  age by moving outwards \citep[e.g.,][]{Katz-stone_1997, Laing_2013,Heesen_2018}. At the center, the radio source has a spectral index $\alpha=-0.30\pm0.05$, which is in agreement with the flat emission reported in the literature \citep[][]{Giovannini_1987} and consistent with the emission from the nucleus of the radio galaxy. The inner plumes are fairly symmetrical with a progressive steepening down to $\alpha\simeq-0.8$. The spectral index steepens across the width of the plumes. The most noticeable differences are in the outer plumes. Whereas in the southern one the spectral index steepens smoothly down to $\alpha\simeq-1.8$, in the northern one the pattern is more varied. We observe a sudden flattening with values down to -0.3 in between the jet and JO36, a flat spectrum in correspondence of JO36 \citep[$\alpha\simeq-0.5$, as expected in star-forming galaxies at low frequencies, e.g.,][]{Heesen_2022}, most likely due to the galactic emission. In the wake of JO36, where we have the most significant uncertainties on the spectral index ($\sim0.2-0.4$), we observe a tentative steepening from -$0.7\pm0.2$ to -1.5$\pm0.3$.

Concerning the Mandolin, the spectral index maps reveal that in the `headstock' (A1) there is a tentative steepening trend along the north-south direction from $\alpha=-1$ to $\alpha=-1.5$ in the southern edge.
\section{Discussion}
In the following, we present a series of analyses and considerations with the purpose of characterizing the JO36-GIN 049 system and addressing whether the two galaxies are interacting.

\subsection{Flux density and spectral index distribution}
\label{profili}
\begin{figure*}
    \centering
    \includegraphics[width=.9\linewidth]{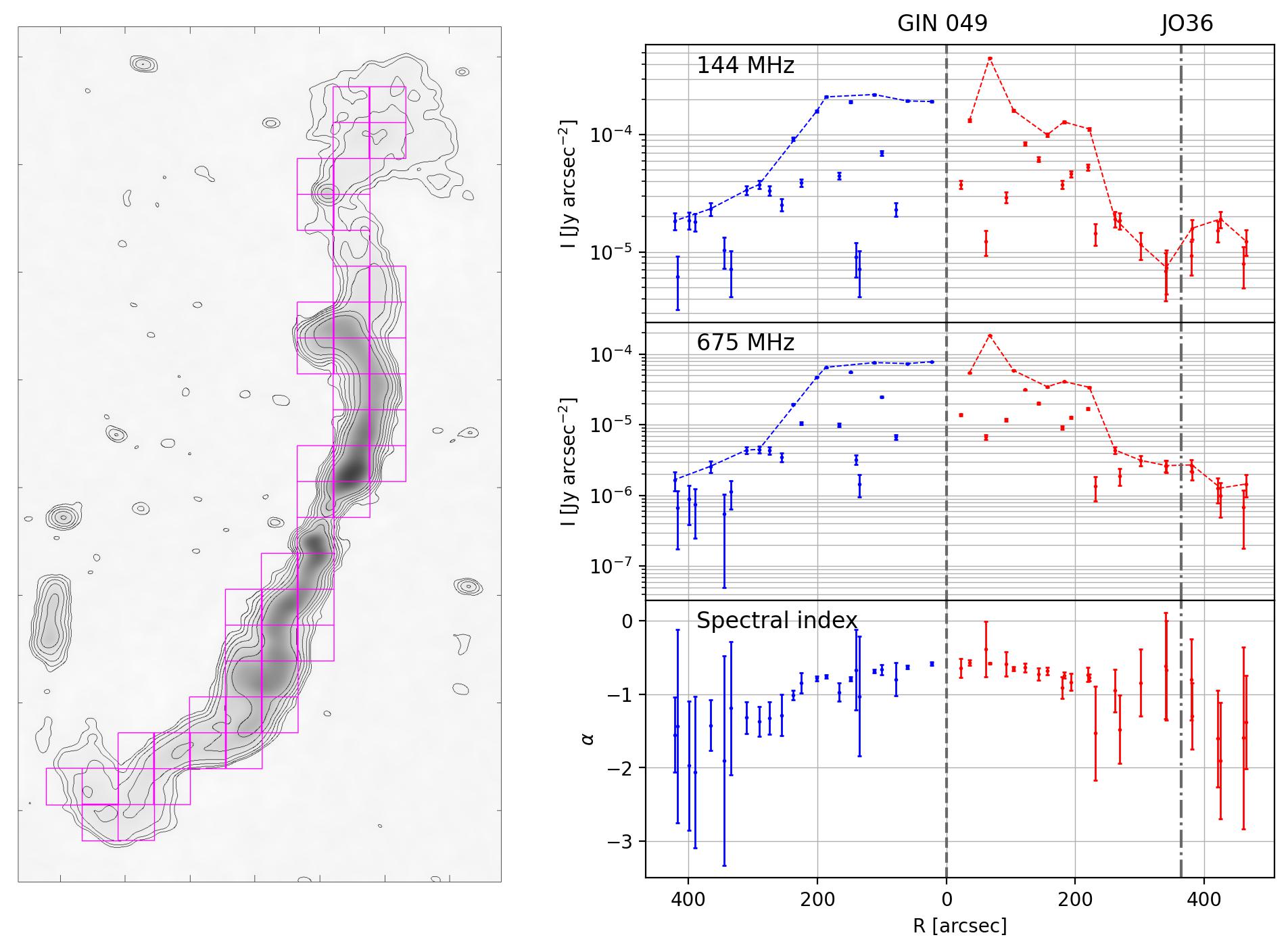}
    \caption{Left: LOFAR image, previously shown in Figure \ref{alfa}, central panels, with on top the 40$\times$40 pixel$^2$ sampling grid; Right: from top to bottom, 144 MHz flux density, 675 MHz flux density, and spectral index vs. the distance from GIN 049. We report northern and southern jets in, respectively, red and blue. The vertical dash-dotted line indicates the position of JO36. }
    \label{sampling}
\end{figure*}

We evaluate the surface brightness and the spectral index throughout the radio source to infer any significant asymmetry between the northern and southern plume which could be attributed to an interaction with JO36. To begin with, we sample the radio emission in the 144 and 675 MHz images with an angular resolution of $17\times17$ arcsec$^2$, which are those used to produce the spectral index map shown in Figure \ref{alfa-image} (central panel), above the 2$\sigma$ with a grid composed by $40\times40$ arcsec$^{2}$ cells with the PT-REX\footnote{\url{https://github.com/AIgnesti/PT-REX}} code \citep[][]{Ignesti_2022}. In each cell where both emissions are above the threshold, we measure the 144 and 675 MHz surface brightness and estimate the corresponding spectral index, and compute the angular distance from the center of GIN 049. The surface brightness errors, and the corresponding spectral index uncertainties, are computed by measuring the RMS contribution in each cell. The results are shown in Figure \ref{sampling}, in which we compare the values for the northern (red) and southern (blue) plumes. For reference, we show the profile along the `ridge lines', which are the regions with the highest surface brightness. 

The distributions are fairly symmetrical within 200 arcsec ($\sim 176$ kpc, projected) from the galaxy, which corresponds to the inner plumes. The ridge profiles indicate that in the northern plume the emission declines more sharply between 200 and 300 arcsec than in the southern one, especially at 144 MHz. The surface brightness profile is fluctuating, also due to the contribution of JO36's radio emission, although the surface brightness at the plume ends is similar. The northern plume appears to be slightly more extended than the southern one. Correspondingly, in the spectral index profile, we observe a tentative flattening coincident with JO36. The extreme flat spectrum emission observed toward the south of JO36 (Figure \ref{alfa-image}) corresponds to the regions in which the 144 MHz surface brightness reaches its minimum, whereas the 675 MHz emission remains similar to the southern counterpart. Therefore, we suggest that the spectral flattening could be a consequence of the uncertain measure of the faint 675 MHz surface brightness, or the result of a complex physical process that favors the escape of the low-energy cosmic rays, thus resulting in a 144 MHz flux density loss. Only future radio observations will be able to confirm the presence of this peculiar trend. Finally, the two outer plumes show a similar spectral index and surface brightness toward their ends.

\subsection{Radio tail morphology}
\label{tail-shape}
Although radio tails at 144 MHz with a characteristic length of $10-40$ kpc have been commonly observed in jellyfish galaxies \citep[][]{Roberts_2021a,Roberts_2021b,Roberts_2021c,Ignesti_2023}, in the case of JO36 the `tail-like' structure (Figure \ref{comparison}, bottom-left panel) shows exceptional characteristics, such as a projected length of $\sim1.9$ arcmin, which corresponds to $\sim140$ kpc at the cluster redshift, and a lack of optical, HI, or X-ray counterparts, at odds with what is observed in other ram pressure stripped galaxies \citep[][]{Ignesti_2021}. Moreover, at odds with what is observed in other ram-pressure-stripped radio tails \citep[][]{Murphy2009,Vollmer_2013,Ignesti_2023}, the `tail' of JO36 becomes wider and brighter with the distance from the stellar disk. The tail-like feature highlighted by the new LOFAR image may confirm that the relative motion of the galaxy and the plume is directed toward the south-east, thus addressing the direction of the galaxy's motion which was one of the open questions left from previous studies (see Section \ref{jo36}).

\subsection{Radio luminosity-SFR}
\label{analisi_sfr}
\begin{figure}
    \centering
    \includegraphics[width=\linewidth]{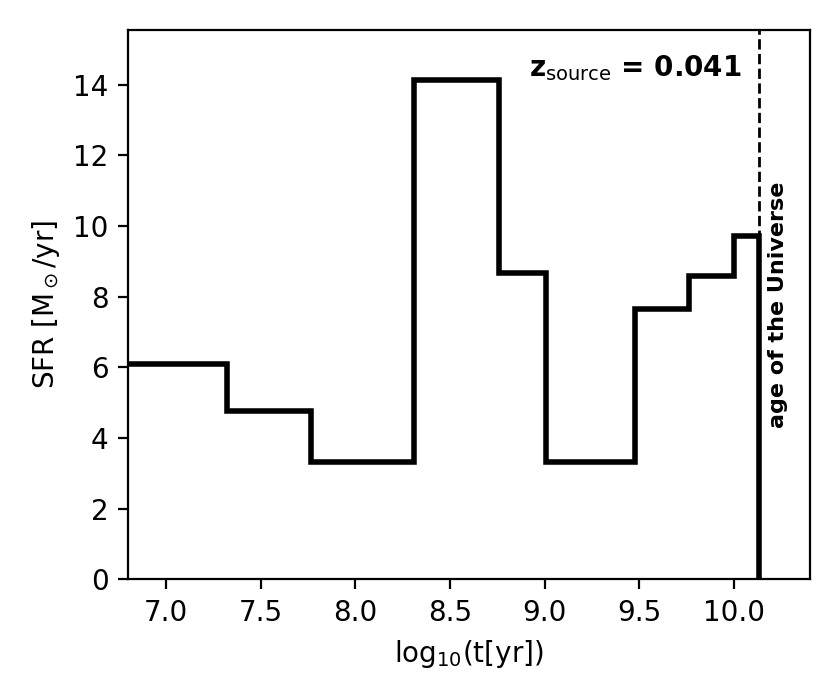}
    \caption{Star formation history of JO36 derived by SINOPSIS from \citet[][]{Fritz2017}.}
    \label{sfh}
\end{figure}

\begin{figure}
    \centering
    \includegraphics[width=\linewidth]{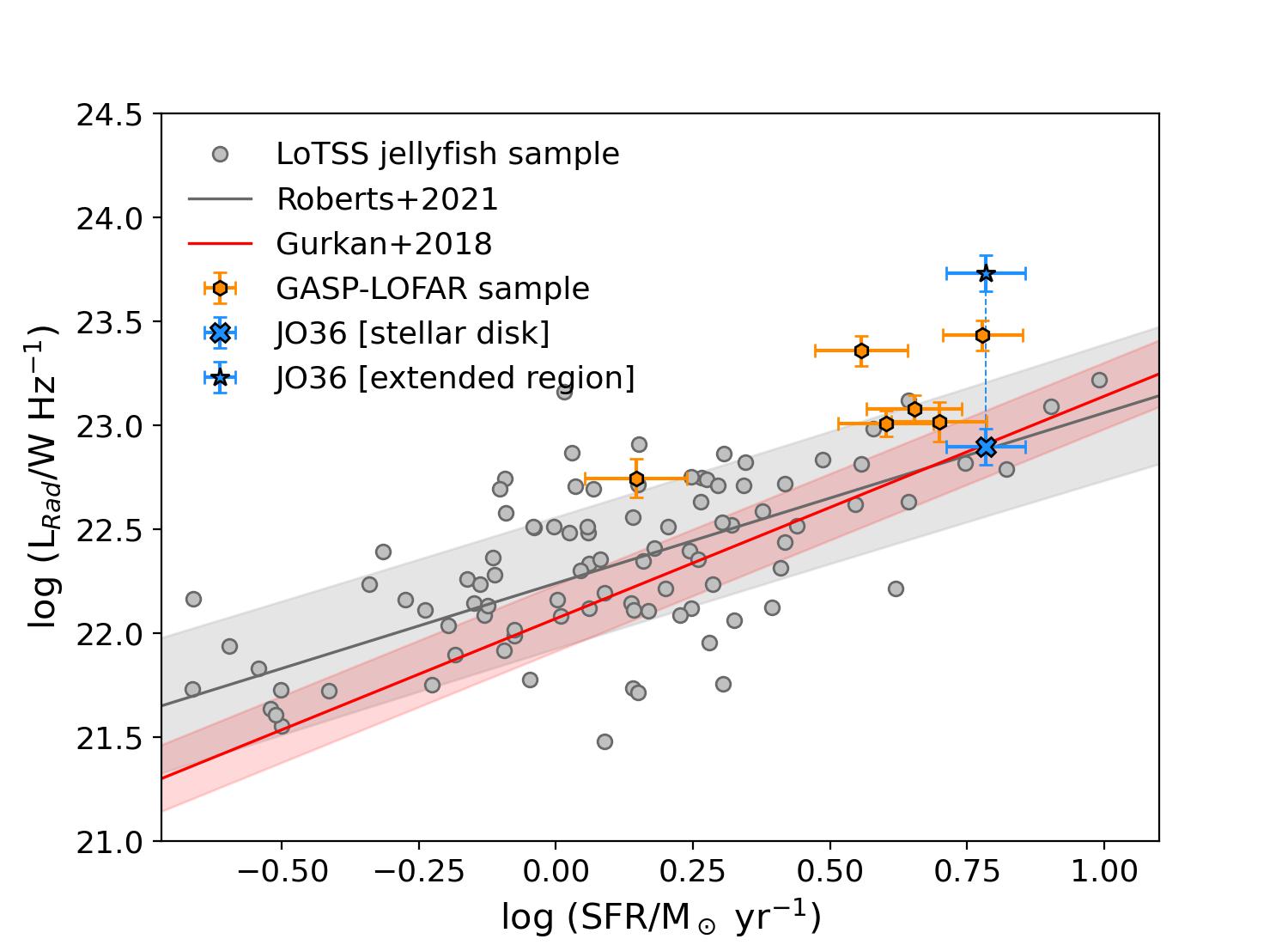}
    \caption{Radio luminosity at 144 MHz vs. SFR showing JO36 stellar disk (blue cross point) and extended region (blue star point), the GASP-LOFAR \citep[gold, from ][]{Ignesti_2022b}, and LoTSS \citep[silver, from ][]{Roberts_2021a} samples. We report the empirical relations between $L_{\text{Rad}}$ and SFR presented by \citet[][]{Gurkan_2018} (red) and \citet[][]{Roberts_2021a} (silver), the latter is fitted on the LoTSS sample. The shaded regions show the dispersion at 1$\sigma$ around the corresponding best-fit relations.}
    \label{sfr-lum}
\end{figure}

In order to address if the `radio tail' actually comes from JO36, we consider the 144 MHz radio luminosity-SFR relation. Star-forming galaxies follow a well-defined relation due to the fact that the radio-emitting relativistic electrons are accelerated in supernovae explosions \citep[see][for a review]{Condon_1992}. Generally, cluster galaxies show a large scatter with respect to the best-fit relation, with jellyfish galaxies having a systematic excess \citep[for further details on this, see][]{Ignesti_2022d}. JO36 has a SFR of $6\pm1$ $M_\odot$ yr$^{-1}$, based on the $H\alpha$ luminosity, and a stellar mass of $6.5\times10^{10}$ $M_\odot$ \citep[][]{Vulcani2018}, which implies that the galaxy is above the SFR-stellar mass main sequence by a factor $\sim3$, hence that it is currently forming stars at a higher rate than that expected based on its stellar mass \citep[][]{Vulcani2018}. Moreover, \citet[][]{Fritz2017} observed that JO36 was subject to a mild burst of star formation in the previous $2-5\times10^8$ yrs (Figure \ref{sfh}, adapted from \citealt[][]{Fritz2017}). Thus we argue that the current SFR excess could be a consequence of this previous burst. We compare the current SFR inferred from the H$\alpha$ luminosity with the 144 MHz radio luminosity measured in the two regions defined in Section \ref{radio_res}, which are the stellar disk and the extended region. In Figure \ref{sfr-lum} we compare JO36 with the other GASP-LOFAR galaxies \citep[gold points, including both the emission in the stellar disks and the extraplanar regions, see][]{Ignesti_2022d}, the LoTSS jellyfish galaxies \citep[][]{Roberts_2021a} (grey points), and the empirical relation at 144 MHz inferred by \citet[][]{Gurkan_2018} (red line). The emission of the stellar disk alone (lower blue cross) is perfectly in agreement with the expected trends, whereas the extended region (upper blue star) is in excess of about one order of magnitude with respect to the empirical trends, and it would be a factor $\sim3$ higher than the most luminous GASP galaxy, JO85, thus making it the most luminous ram pressure stripped galaxy observed by LOFAR.   

The combined analysis of $L_R$-SFR relation and the SFH strongly indicate separate origins for the radio emission in the disc of JO36 and in the extended region. First, the stellar disk emission follows the best-fit relations, thus suggesting that either no external process has enhanced the radio luminosity \citep[e.g.,][]{Gavazzi2006, Murphy_2009b}, or that the star formation has been uniform for, at least, the synchrotron radiative time \citep[$<$200 Myr at 144 MHz, e.g., ][]{Basu_2015, Ignesti_2022d}. This latter hypothesis can be further tested by comparing the radio and infrared luminosity. In the stellar disk of JO36, \citet[][]{Fritz2017} reported a flux density of 0.47 Jy at 250 $\mu$m measured with \textit{Herschel}, which entails a luminosity of $2\times10^{24}$ W Hz$^{-1}$. By applying the empirical radio-infrared relation presented by \citet[][]{Gurkan_2018} (Equation 4), the corresponding expected radio luminosity would be $8\times10^{22}$ W Hz$^{-1}$, which is in agreement with the observed luminosity of $7.9\times10^{22}$ W Hz$^{-1}$. Therefore, the agreement between the radio, infrared, and H$\alpha$ emission confirms that the SFR in the stellar disk has been almost constant in the last $\sim10^8$ yr. 

Consequently, a factor $\sim10$ radio excess, which is what we measure in the combined extended region (Figure \ref{sfr-lum}), would be possible only if the SFR averaged over the timescale of the synchrotron radiative time ($<10^8$ yrs) was a factor 10 larger than that measured within the H$\alpha$-emitting timescales \citep[$\sim10$ Myr, see][]{Kennicutt_2012}, which is incompatible with the derived SFH. Therefore, we suggest that the `echoes of the past star formation', which is the apparent excess in radio luminosity resulting from the SFR decreasing with a time-scale that is shorter than the relativistic electrons radiative time \citep[][]{Ignesti_2022d}, cannot be responsible for the extended region radio excess, and we conclude that radio plasma in the extended region does not come from JO36, but is instead the result of the interaction with the radio plume.

\subsection{Truncation radius}
\label{analisis_trunc}
\begin{figure*}
    \includegraphics[width=0.5\linewidth]{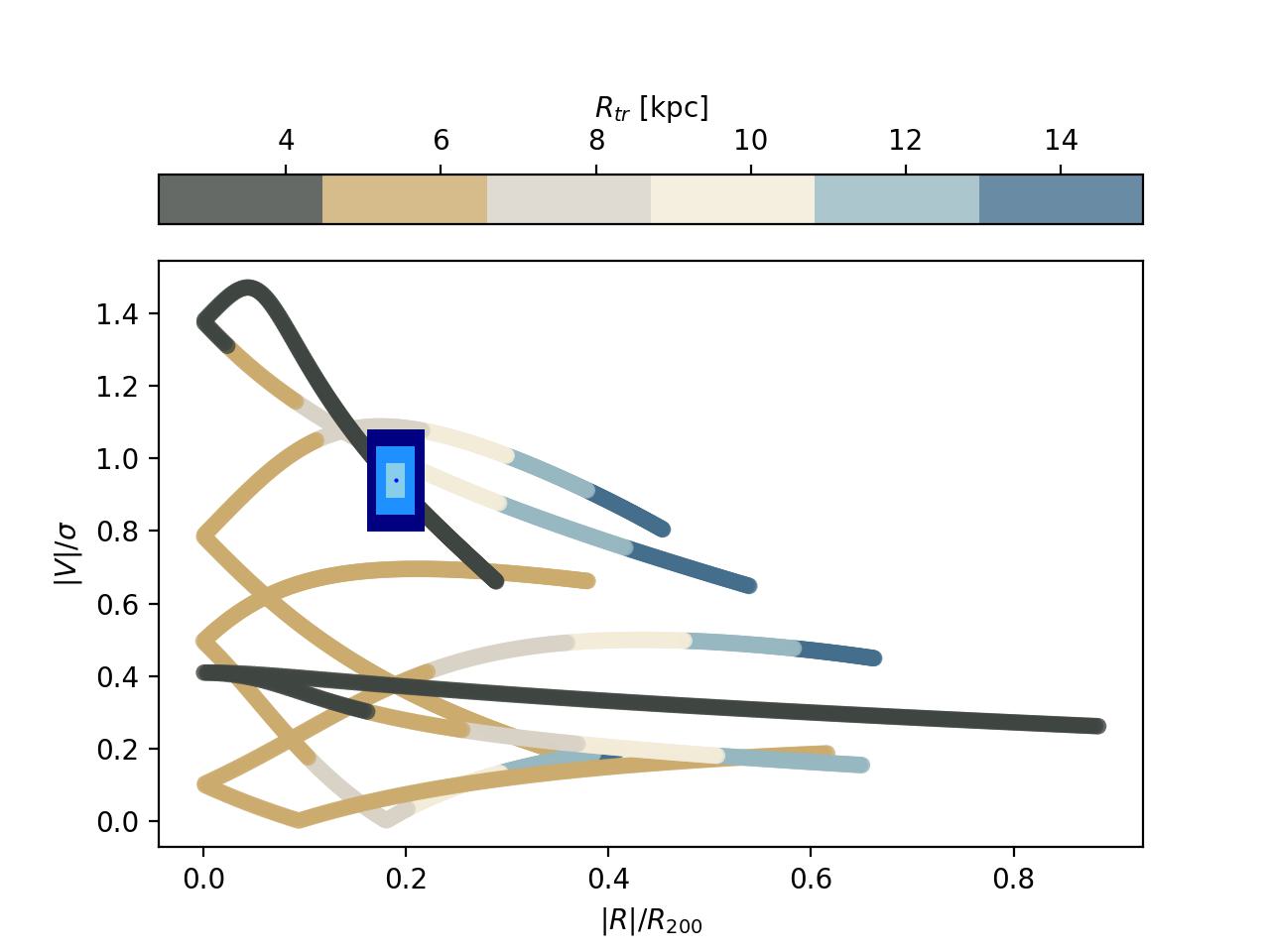}
    \includegraphics[width=0.5\linewidth]{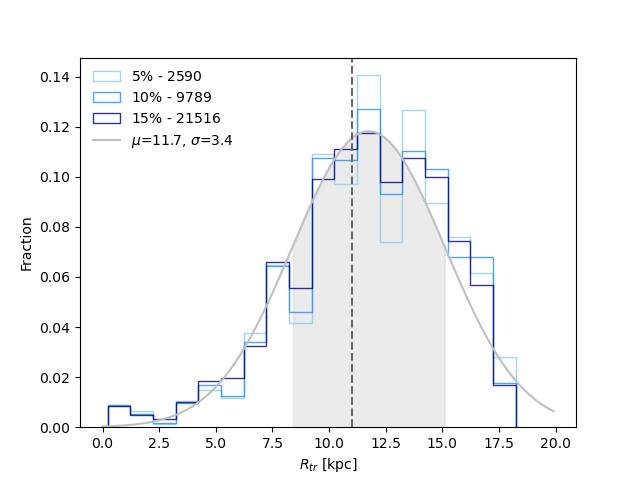}
    \caption{\label{trunc}Left: example of 5 random trajectories on the phase-space diagram, color-coded for the value of $R_{tr}$, and the 5, 10, 15$\%$ scatter regions around the phase-space coordinates of JO36; Right: resulting normalized distributions with a Gaussian fit. The vertical dashed line corresponds to the observed radius, and the grey shaded area indicates the Gaussian fit 1$\sigma$ confidence region. In the legend are reported the number of solutions of each distribution, and Gaussian centroid, $\mu$, and dispersion, $\sigma$.}
\end{figure*}

Finally, we investigate for signatures of a potential interaction between the radio plume and JO36 by studying the morphology of the latter, specifically on the resulting truncation due to ICM ram pressure \citep[][]{Gunn1972}. Specifically, we test if the H$\alpha$ truncation radius \citep[11 kpc,][]{Fritz2017} is consistent with the ICM ram pressure by running a numerical simulation. Hereafter we summarize the analysis, while a more detailed description is presented in Appendix \ref{appendix_truncation}. First, we model a spiral galaxy with baryonic and dark matter properties compatible with JO36, moving within the gravitational potential of a (spherical) A160-like cluster and subject to ram pressure stripping from the ICM. Then we simulate 10000 infalling orbits and track each of them for 3 Gyr. Each orbit starts from a random point on a spherical shell with radius equal to the virial radius of the cluster \citep[$R_{200}=1.6$ Mpc,][]{Gullieuszik_2020}, and has a starting velocity defined by the two components, radial and transverse, whose values are randomly selected from cosmologically-motivated probability distribution (see Appendix \ref{appendix_truncation}). For each orbit, we compute the evolution of the truncation radius by assuming a face-on stripping (i.e., a wind  perpendicular with respect to the disk), thus neglecting the effects induced by different wind-angle configurations \citep[e.g.,][]{Roediger_2006, Jachym_2009, Bekki_2014,Steinhauser2016, Farber_2022, Akerman_2023}. Then we project the orbit along a random line-of-sight to derive the corresponding trajectory in the phase-space diagram (Figure \ref{trunc}, left panel). We note that, by including all positions along the orbit, we are inherently assuming a constant rate of galaxy infall into the cluster. Finally, in the cumulative phase-space composed of all orbits, we collect the truncation radii of all galaxies with phase-space coordinates within the 5, 10, 15$\%$ scatter of the values observed for JO36 ($|R|/R_{200}=$0.19, $|V|/\sigma=$0.94).

The three resulting distributions, shown in Figure \ref{trunc} (right panel), are similar and are centered around $\sim11$ kpc with a scatter of a few kpc and tails toward low values, representing the solutions found after the pericenter passage. For reference, we fit the entire distribution obtained within the $10\%$ scatter with a Gaussian finding $\mu=11.7$ kpc and $\sigma=3.4$ kpc, which encompasses the observed truncation radius of 11 kpc. Therefore, we conclude that the stripping of JO36 is consistent with the action of the thermal ICM and, thus, the truncation radius cannot discriminate the potential role played by other physical processes induced by the passage through a magnetized, relativistic plasma.

\subsection{The emerging picture}
As the main properties of the anomalous radio tail associated with JO36 are markedly different from those routinely observed in ram-pressure stripped cluster galaxies (Sections \ref{tail-shape}, \ref{analisi_sfr}), one may wonder whether alternative explanations for its origin are feasible. One possibility is that the peculiar morphology could result from an accidental overlap of the northern plume and JO36 being located in the background/foreground with respect to GIN 049. In this case, the extended `tail-like' feature would be the extremely fortunate result of the overlap between the galaxy and a substructure in the northern plume. However, it is unclear how the sole projection effect could produce the coherent `radio tail' highlighted by the UV-cut (Figure \ref{panel}, bottom-left panel), as well as the spectral index trend in the wake of JO36. Therefore, on the basis of the radio morphology and the peculiar star formation history of JO36, we propose that the galaxy has actually interacted with the radio plume of GIN 049. The radio morphology seems to indicate that the JO36 is moving along the NW-SE direction, albeit the presence of the isolated H$\alpha$ blobs may suggest that a significant component of the motion is along the line-of-sight \citep[][]{Fritz2017}. The analysis of the northern plume properties (Section \ref{profili}) and H$\alpha$ truncation (Section \ref{analisis_trunc}) indicate that the interaction was `mild', in the sense that it did not leave any significant signature in terms of relativistic electrons re-acceleration or alterations in the galaxy morphology. At the same time, the spectral steepening trend observed in the wake of JO36 (Figure \ref{alfa-image}, bottom panels) suggests the presence of a shock associated with the galaxy that has advected brighter and flatter-spectrum material up the plume. Therefore we argue that the galaxy has crossed the plume and hindered the flow of the plume. If the combined speed of the flow of material up the plume and the bulk motion of the galaxy is mildly supersonic with respect to the local sound speed, which is unknown, or the plume flow is internally trans-sonic and the relative velocity of JO36 and GIN049 is much less than the plume velocity (i.e., JO36 acts as a `stationary obstacle' to the plume flow), then a shock is expected to form when the plume passes the galaxy.

The peculiar morphology of JO36's extended emission could be explained by the so-called `magnetic draping', that is the accretion of an external, ordered magnetic field on the top of an object moving in a magnetized medium with a velocity that is higher than the local Alfven velocity \citep[e.g.,][]{Dursi_2008,Pfrommer_2010, Muller_2021, deGasperin_2022}. The physical conditions inside the northern plume, which determine the local Alfven velocity, are unknown. Nevertheless, for a relativistic plasma with adiabatic index $\gamma=4/3$ the ratio between the speed of sound, $c_s$, and Alfven velocity, $V_A$, is:
\begin{equation}
\frac{c_s}{V_A}=\frac{\sqrt{\gamma\frac{P}{\rho}}}{\frac{B}{\sqrt{\mu_0 \rho}}}=\sqrt{\frac{4}{3}\frac{\mu_0 P}{B^2}}=\sqrt{\frac{2}{3}\beta},
\end{equation}
where $P$, $\rho$, and $B$ are, respectively, the local pressure, mass density, and magnetic field, $\mu_0$ is the vacuum permeability, and $\beta=2\mu_0 P/B^2$ is the so-called `plasma-$\beta$'. Therefore for $\beta>1.5$, $c_s>V_A$. Numerical simulations of radio galaxies suggest that in the plumes the physical conditions entail $\beta\gg 1$ \citep[e.g.,][]{Jones_2002, Nolting_2019,Giri_2022}, so
if JO36 is moving super-sonically ($V>c_s$) then it is also super-Alfvenic ($V>c_s>V_A$), and, thus, crossing the magnetized plume can produce a `magnetic drape'. In this framework, we speculate that the long tail-like feature might be the result of the galaxy `dragging away' the magnetic field from the plume. However, we note that the resulting emission differs from the `tadpole-like' shape predicted by simulations \citep[][]{Dursi_2008,Muller_2021}, as the `tail' appears to widen with distance and we do not observe any enhanced radio emission ahead of JO36. We suggest that these discrepancies could either be due to projection effects that blend the drape emission with the northern plume, or due to the fact that the proposed interaction is more complex than those currently explored by numerical simulations.

The magnetic drape is also supposed to be able to `shield' the galaxy from the ICM by damping the thermal conduction and, consequently, the mixing between the ICM and ISM \citep[][]{Muller_2021, Campitiello_2021, Bartolini_2022}. In this case, these effects appear to be negligible as the truncation radius appears to be consistent with the thermal ICM ram pressure (Section \ref{analisis_trunc}), and we do not observe any H$\alpha$ cloud `preserved' by the putative drape (Figure \ref{halpha}). The latter piece of evidence could be explained if the galaxy passed through the plume when it was already in an advanced phase of stripping, hence when it was already without a tail of ionized ISM. 

Finally, we can estimate a tentative dynamical time for the galaxy to cross the plume:
\begin{equation}
    \tau=\frac{L}{\sigma}\simeq265\left( \frac{L}{\text{200 kpc}}\right)\left(\frac{V}{738\text{ km s}^{-1}} \right)^{-1}\text{ Myr,}
\end{equation}
where $L=200$ kpc is roughly the outer plume diameter (Figure \ref{comparison}, top-left panel). This dynamical age would be consistent with the star formation burst lookback time (Section \ref{analisi_sfr}, Figure \ref{sfh}), thus suggesting that the SFR burst might have been induced by the first impact of JO36 on the radio lobe, perhaps due to entering in a higher-pressure environment or being traversed by the shock induced by the super-sonic flow of the plume. 

The magnetic draping scenario provides us with a striking prediction that can be tested with future observations. The large-scale, ordered magnetic field in the wake of JO36, resulting from the passage through the magnetized lobe, would result in highly-polarized radio emission aligned with the galaxy motion \citep[as observed in another ram pressure-stripped galaxy, see][]{Muller_2021}. Furthermore, the presence of a magnetized component embedded in the plume emission can be tested by means of the Faraday rotation analysis. Therefore, a deep observation at frequencies higher than 1 GHz, where the polarized emission is more visible, is now desirable to test this scenario.

\subsection{A comparison with similar systems}
We can draw a comparison between the JO36-GIN 049 system and other similar, puzzling systems, where the interaction between AGN jets and external galaxies is possibly observed with consequent positive feedback induced. Among these there are  Minkowski's object \citep[e.g.,][]{Minkowski_1958, vanBreugel_1985, Croft_2006,Salome_2015,Zovaro_2020} in NGC541, and the galaxy-jet interaction in FRII radio galaxies, such as the cases of 3C 441 \citep[][]{Lacy_1998} and 3C 285 \citep[][]{vanBreugel_1985}. Finally, a potential lobe-galaxy interaction has been suggested to explain the complex morphology of NGC 326 \citep[][]{Hardcastle_2019} and RAD12 \citep[][]{Hota_2022}.

First, we highlight the differences. To begin with, our studies suggest that the JO36 encounter was moderately energetic, whereas in the other cases, the observations suggested that the galaxies had been hit by the relativistic jet. Moreover, the other galaxies are irregular with peculiar properties in the optical spectrum. However, all these systems, including JO36, share a common trait, which is a star formation burst/enhancement event in the past. These events seem to be related to `positive feedback' from the interaction with their respective radio galaxies, either due to a direct compression exerted by the jets or the passage through a high-pressure medium. This physical phenomenon can affect multiple galaxies in dense environments, resulting in a global SFR enhancement \citep[][]{Gilli_2019}. Addressing the exact physical processes that drive these events is complex and beyond the scope of our work. Nevertheless, the observations reported in this article can help to drive future studies aimed in this direction. For instance, the bimodality of regular/irregular galaxies, associated with the encounter with remnant plasma or a relativistic jet, might outline that the relativistic plasma energy determines the way in which the encounters affect the galaxy morphology.

\section{Conclusions}
We presented brand new LOFAR and uGMRT images of a peculiar interaction between two galaxies, namely JO36, a ram-pressure stripped spiral, and GIN 049, the FRI BCG of Abell 160. The new LOFAR data revealed that GIN 049 is a giant radio galaxy, with a projected physical size of, at least, 744 kpc, and that JO36 is embedded in the northern radio plume. The new radio images at 144 and 675 MHz allowed us to produce new high-resolution spectral index maps.

We investigated the nature of the JO36-GIN 049 interaction and its implications by studying the properties of the galaxy and the radio plume. JO36 has an integrated radio luminosity of  $5.4\times10^{23}$ W Hz$^{-1}$ and a giant radio tail with a projected physical size of $\sim150$ kpc. Both these properties exceed what has been commonly observed in ram pressure-stripped galaxies, and we conclude that they are the result of the interaction with the plume.  In turn, the first impact of the galaxy on the radio plume might have triggered a mild SFR enhancement in the past $\sim2-5\times10^8$ yrs, in agreement with the SFH of JO36 derived from the modeling of the MUSE spectra. Concerning GIN 049, the northern and southern lobes show similar surface brightness and spectral index, but they differ in morphology, so we conclude that the JO36 encounter did not result in a detectable re-acceleration of the old radio plasma. However, the galaxy could have re-shaped the northern plume by dragging the lobe magnetic field along its orbit due to the magnetic draping. Future polarimetric observations at frequencies higher than 1 GHz can test this speculative scenario by confirming the presence of a large-scale ordered magnetic field (i.e., highly polarized radio emission) elongated along the galaxy wake, which is the characteristic signature of magnetic draping. This system represents a new, unique laboratory to study the astrophysics of relativistic plasmas and will be the object of future studies.

\section*{Acknowledgments}
We thank the Referee for the constructive report that improved the presentation of this study. A.I. thanks F. Vazza for the useful discussion and A. Acuto for suggesting the name `Mandolin'. This work is the fruit of the collaboration between GASP and the LOFAR Survey Key Project team (``MoU: Exploring the low-frequency side of jellyfish galaxies with LOFAR", PI A. Ignesti).  A.I. acknowledges the INAF founding program 'Ricerca Fondamentale 2022' (PI A. Ignesti). M.B. acknowledges financial support from the agreement ASI-INAF n. 2017-14-H.O and from the PRIN MIUR 2017PH3WAT “Blackout”. B.V., R.P., M.G. acknowledge the Italian PRIN-Miur 2017 (PI A. Cimatti). J.F. acknowledges support from DGAPA PAPIIT project IN110723. A.D. acknowledges support by the BMBF Verbundforschung under the grant 05A20STA. A.I. thanks the music of Black Sabbath for providing inspiration during the preparation of the draft.

Based on observations collected at the European Organization for Astronomical Research in the Southern Hemisphere under ESO programme 196.B-0578. This project has received funding from the European Research Council (ERC) under the European Union's Horizon 2020 research and innovation programme (grant agreement No. 833824). LOFAR \citep[][]{vanHaarlem_2013} is the Low Frequency Array designed and constructed by ASTRON. It has observing, data processing, and data storage facilities in several countries, which are owned by various parties (each with their own funding sources), and that are collectively operated by the ILT foundation under a joint cientific policy. The ILT resources
have benefited from the following recent major funding sources: CNRS-INSU, Observatoire de Paris and Université d'Orléans, France; BMBF, MIWF-NRW, MPG, Germany; Science Foundation Ireland (SFI), Department of Business, Enterprise and Innovation (DBEI), Ireland; NWO, The Netherlands; The Science and Technology Facilities Council, UK; Ministry of
Science and Higher Education, Poland; The Istituto Nazionale di Astrofisica (INAF), Italy. This research made use of the Dutch national e-infrastructure with support of the SURF Cooperative (e-infra 180169) and the LOFAR e-infra group. The Jülich LOFAR Long Term Archive and the German LOFAR network are both coordinated and operated by the Jülich Supercomputing Centre (JSC), and computing resources on the supercomputer JUWELS at JSC were provided by the Gauss Centre for Supercomputing e.V. (grant CHTB00) through the John von Neumann Institute for Computing (NIC). This research made use of the University of Hertfordshire high-performance computing facility and the LOFAR-UK computing facility located at the University of Hertfordshire and supported
by STFC [ST/P000096/1], and of the Italian LOFAR IT computing infrastructure supported and operated by INAF, and by the Physics Department of Turin university (under an agreement with Consorzio Interuniversitario per la Fisica Spaziale) at the C3S  Supercomputing Centre, Italy. We thank the staff of the GMRT that made these observations possible. GMRT is run by the National Centre for Radio Astrophysics of the Tata Institute of Fundamental Research. This research made use of Astropy, a community-developed core Python package for Astronomy \citep[][]{astropy_2013, astropy_2018}, and APLpy, an open-source plotting package for Python \citep[][]{Robitaille_2012}.

\bibliography{sample631}{}
\bibliographystyle{aasjournal}
\appendix
\section{{Truncation radius estimate}}
\label{appendix_truncation}
\subsection{Cluster properties}
\begin{figure}[b]
    \centering
    \includegraphics[width=0.6\linewidth]{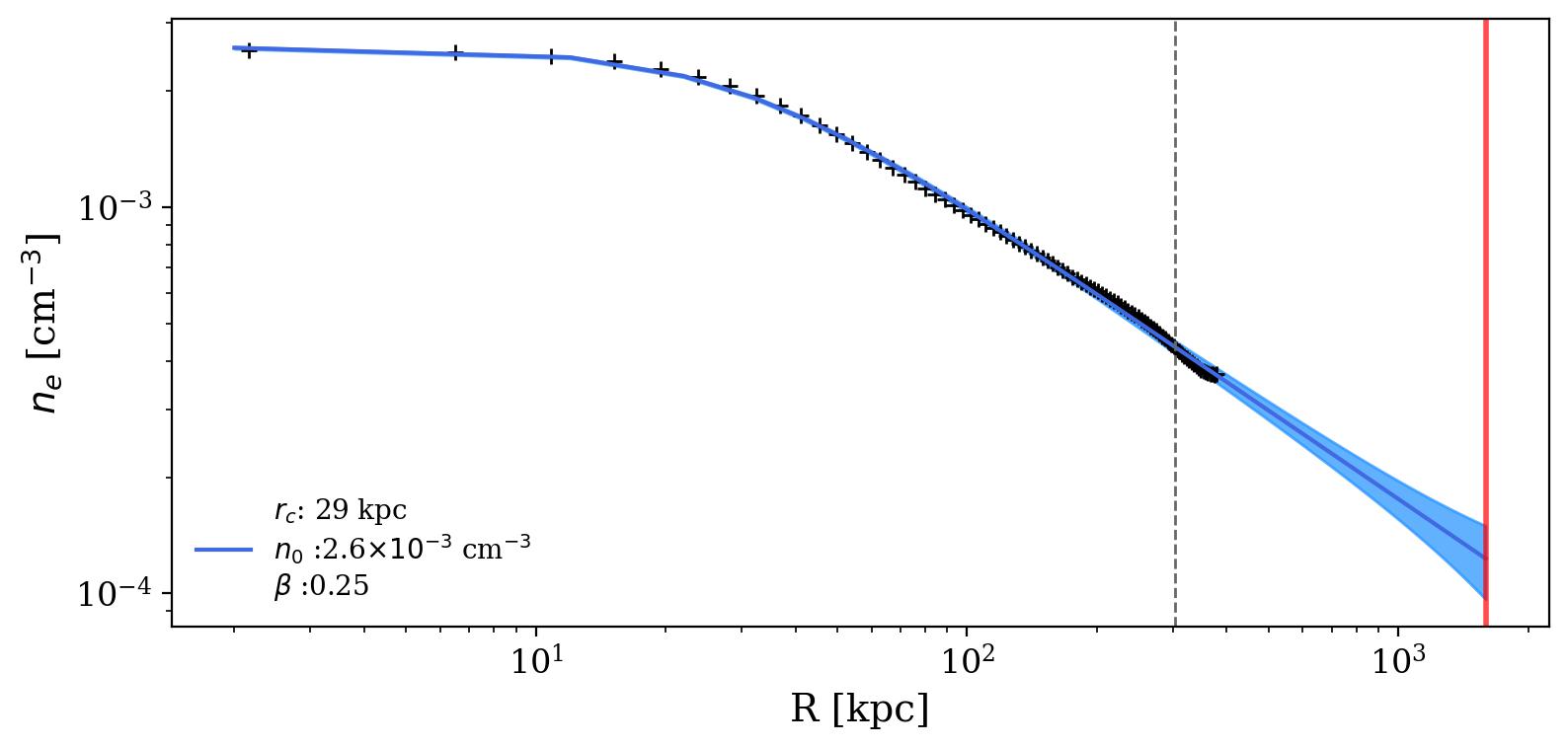}
    \caption{ACCEPT $n_e$ profile with the best-fit $\beta$ model. The blue shaded area indicates the $95\%$ confidence interval. The grey-dashed and red vertical lines point out the JO36 projected clustercentric distance and $R_{200}$, respectively.}
    \label{betamodel}
\end{figure}

In order to build a model for the extended mass profile of the A160 cluster, we assume the ICM of the cluster is isothermal and in hydrostatic equilibrium with the potential of the cluster. We base the ICM model on the properties reported in the ACCEPT survey \citep[][]{Cavagnolo_2009}. The thermal properties are measured from the \textit{Chandra} X-ray observation 3219 (exposure time 58.5 ks). We make use of the average ICM temperature (2.7 keV) and the electron density $n_e$ profile to model the ICM distribution. Specifically, under the assumption of isothermality, we fit the observed $n_e$ profile with a $\beta$ model:
\begin{equation}
    n_e(r)=\left[1+\left(\frac{r}{r_c} \right)^2 \right]^{-\frac{3}{2}\beta},
\end{equation}
where $n_0$ is the central density, $r_c$ the core radius, and $\beta=\mu m_p \sigma^2/kT$ \citep[][]{Cavaliere-Fusco_1976}. We measure $r_c=29$ kpc, $n_0=2.6\times10^{-3}$ cm$^{-3}$ and $\beta=0.26$, the best-fit profile is shown in Figure \ref{betamodel}. However, we note that there are a series of caveats in our fit. To begin with, the observation is not deep enough to map the ICM thermal emission up to $R_{200}$, and, thus, our fit is based on the emission up to $\sim300$ kpc. Therefore, the predicted ICM density in the cluster periphery is an extrapolation and it may underestimate the real ICM density. Second, the cluster morphology is complex and not spherical \citep[][]{Cavagnolo_2009}, as is instead assumed by the $\beta$ model, thus resulting in a relatively flat $\beta$ index.  
\subsection{Galaxy properties}
The mass model for JO36 is based on the H$\alpha$ rotation curve, which we derive from the MUSE data using the following approach.
We first produce a continuum-subtracted H$\alpha$ velocity-cube, made by all MUSE spectral channels within $\pm600$ km\,s$^{-1}$ from the center of the H$\alpha$ line. 
Continuum subtraction is tailored to a region around the H$\alpha$ line, and is performed spaxel-by-spaxel by fitting the spectrum in a $120\AA$-wide window from the peak of the H$\alpha$ emission with a first-order polynomial, after the masking of the relevant emission lines in that window (H$\alpha$, [NII]$\lambda6548$, [NII]$\lambda6584$).
We tested the use of higher-order polynomials for the continuum subtraction, finding similar results.

\begin{figure}
\begin{center}
\includegraphics[width=0.52\textwidth]{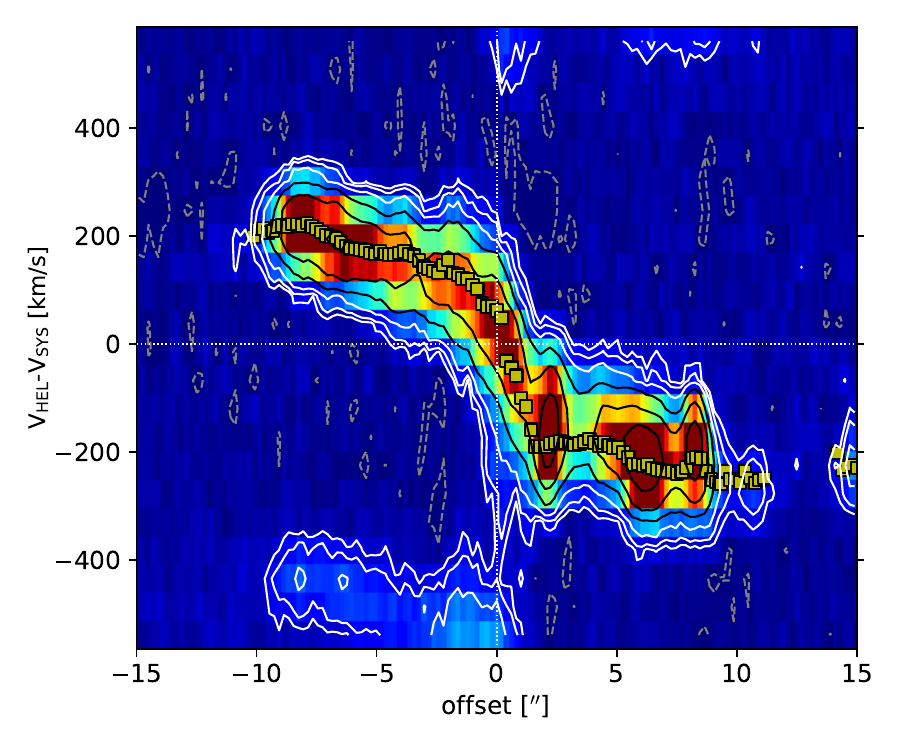}
\includegraphics[width=0.47\textwidth]{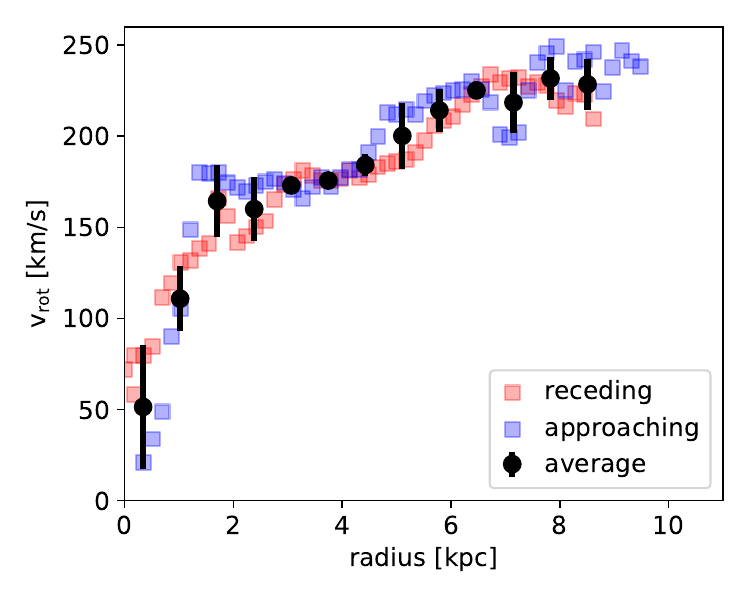}
\caption{H$\alpha$ kinematics of JO36. \emph{Left:} position-velocity slice along the galaxy's major axis, extracted from the MUSE data. The yellow squares show the rotation curve determined with the envelope tracing method. The bright regions at velocities $|v|>400$ km\,s$^{-1}$ are due to the [NII] doublet entering the velocity range of the datacube. Intensity contours are drawn at $3\times2^N$ ($N\geq0$) times the rms-noise (solid lines). A negative contour at $-3$ times the rms-noise is also shown with grey dashed lines. \emph{Right:} the corresponding rotation curve. The approaching and receding sides are shown separately as blue and red squares, respectively. The black circles show the mean rotation velocity. 
}
\label{f:JO36_kinematics}
\end{center}
\end{figure}

The rotation curve extraction from the H$\alpha$ cube exploits the fact that JO36 is an edge-on system \citep{Fritz2017}, and is based on the `envelope-tracing' method \citep[e.g.][]{SancisiAllen79}.
In practice, we extract the position-velocity diagram across the H$\alpha$ kinematic major-axis (shown in the left panel of Fig.\,\ref{f:JO36_kinematics}), and fit the high-velocity tails (i.e., the regions at velocities above the intensity peak) of individual H$\alpha$ profiles with a Gaussian function, whose mean gives the rotation velocity at a given radius.
In our fit we assume a constant velocity dispersion $\sigma$ given by the combination of an instrumental and an intrinsic broadening \citep[e.g.][]{Marasco+23}, $\sigma=(\sigma^2_{\rm MUSE} + \sigma^2_{\text{int}})^{1/2}$, where $\sigma_{\rm MUSE}=50$ km\,s$^{-1}$ and $\sigma_{\text{int}}=30$ km\,s$^{-1}$ \citep[typical of low-$z$ star forming galaxies, e.g.][]{Epinat+10,Green+14}.

The rotation velocity shows substantial (up to $50$ km\,s$^{-1}$) differences between the receding and the approaching sides of the galaxy.
We have adjusted the systemic velocity of JO36 so that the velocity differences between the two sides are minimized.
The right panel of Fig.\,\ref{f:JO36_kinematics} shows the resulting rotation curve for the approaching (blue) and receding (red) sides separately, along with the mean rotation (black circles with error bars) averaged over radial bins of $0.7$ kpc, corresponding to the MUSE spatial resolution.
The outermost 3-4 points of the mean rotation curve show signs of flattening ($v_{\rm flat}\simeq225$ km\,s$^{-1}$), which is relevant to constrain the mass of the dark matter halo.
The shallow rise of the rotation curve indicates the absence of a substantial bulge component, in agreement with the stellar surface density map derived with \textsc{SINOPSIS}. 

A fully three-dimensional construction of the gravitational field about JO36 is constructed, consisting of a spherical NFW dark matter halo \citep[][]{Navarro_1997}, and an exponential disk of gas and stars. For the NFW halo, we assume a virial mass of $1.51\times10^{12}$ $M_{\odot}$, and a concentration parameter of 18. The exponential disk of stars is modeled by combining three Miyamoto-Nagai disks \citep[][]{Miyamoto1975} in the manner described in \citet[][]{Smith2015} in order to produce a total stellar disk mass of $5\times10^{10}$ $M_\odot$ with an exponential scale-length of 4.63 kpc that is razor-thin. The gas exponential disk is also modeled as razor-thin using the same approach. The gas fraction is chosen to be 10\% of the stellar disk mass (a typical gas fraction for a galaxy of this mass), and the gas disk scalelength is 1.7 times larger than the stellar disk scalelength \citep[][]{Cayatte1994}. A bulge component is not included. We measure the circular velocity within the plane of the disk as a function of radius from the galaxy center and find the above model is in good agreement with the observed rotation curve.

\subsection{Simulation}
Galaxy models are initialized at the virial radius of the cluster as they fall into the cluster for the first time. The parameters describing their models are summarized in Table \ref{param_sim}. We choose an initial radial velocity ($V_{\text{rad}}$) and tangential velocity ($V_{\text{tang}}$) component based on measurements made in cosmological simulations as halos first fall into clusters \citep[][]{Smith_2022}. Specifically, the initial values of each orbit are randomly extracted from two distributions, respectively defined by $0.87\pm0.36$ and $0.63\pm0.33$ in units of the cluster velocity dispersion $\sigma=738$ km s$^{-1}$ \citep[][]{Gullieuszik_2020}. Here, we randomly sample $V_{\text{rad}}$ and $V_{\text{tang}}$ from these distributions, such that each individual orbit differs from the others.

The motion of the galaxy along its orbit in the cluster potential is computed using a particle stepper technique. At each time step, the ICM density and galaxy orbital velocity is used to compute the ram pressure. A \citet[][]{Gunn1972} style approach is used to compute the gas truncation radius resulting from that ram pressure. This means the strength of the ram pressure is compared to the restoring gravitational force of the galaxy, and the truncation radius is found at the location where they are equal. However, as noted in \citet[][]{Abadi_1999}, in a three-dimensional galaxy model including a spherical component (the dark matter halo), the maximum restoring force at a fixed cylindrical radius is not always on the plane of the disk. Therefore, at each cylindrical radius in our three-dimensional model, we ascertain the maximum restoring force considering a range of heights above the disk plane, and the truncation radius is calculated using this maximum.
\begin{table}[]
    \centering
    \begin{tabular}{lc}
\toprule
\toprule
Star disk mass [$M_\odot$] & $5.2\times10^{10}$\\		
Gas disk mass [$M_\odot$]& $5.2\times10^{9}$\\		
Star disk scalelength [kpc]& 4.6\\		
gas disk scalelength [kpc]& 7.8\\
ICM beta model, $r_c$ [kpc] &47		\\
ICM average temperature [K] & $3.1\times10^7$\\
ICM beta model $\beta$ & 0.25 \\
ICM beta model, $\rho_0$ [g cm$^3$] & $4.7\times10^{-27}$ 	\\
\midrule
    \end{tabular}
    \caption{Galaxy and cluster model parameters.}
    \label{param_sim}
\end{table}
\end{document}